\documentclass[12pt]{article}
\def\R{{\bf R}}
\def\d{\partial}

\begin{document}

\title{Inextendibility of expanding cosmological models with symmetry}

\author{Mihalis Dafermos\\
University of Cambridge\\
Department of Pure Mathematics and Mathematical Statistics\\
Wilberforce Road, Cambridge CB3 0WB, UK\\and\\
Alan D. Rendall\\
Max Planck Institute for Gravitational Physics\\ 
Albert Einstein Institute\\ Am M\"uhlenberg 1\\
D-14476 Golm, Germany}

\date{}

\maketitle

\begin{abstract}
A new criterion for inextendibility of expanding cosmological models with 
symmetry is presented. It is applied to derive a number of
new results and to simplify the proofs of existing ones. In particular 
it shows that the solutions of the Einstein-Vlasov system with 
$T^2$ symmetry, including the vacuum solutions, are inextendible in
the future. The technique introduced adds a qualitatively new element to 
the available tool-kit for studying strong cosmic censorship.
\end{abstract}

\section{Introduction} 

A central open question in mathematical relativity is that of strong 
cosmic censorship \cite{me, chrusciel}. It is convenient to formulate this in
terms of the maximal Cauchy development \cite{cbgeroch}. Intuitively,
this is the largest globally hyperbolic spacetime satisfying the Einstein
equations which evolves from prescribed initial data on a Cauchy surface.
Implicit in this is the assumption that the Einstein equations are coupled
to some suitable equations describing the motion of matter and that the
matter model chosen does not suffer from unphysical singularities. (See
\cite{sing}, section 4 for a discussion of the latter point.) The strong 
cosmic censorship hypothesis says that the maximal Cauchy development should 
be inextendible. In other words it should not be possible to embed the 
maximal Cauchy development non-trivially in any other regular spacetime, at 
least for generic initial data.

There are some ambiguities in this formulation as just stated. When we
say an extension do we mean only that the geometry can be extended, or
that this should also hold for the matter fields? Should the extended
metric and matter fields satisfy the Einstein equations and the 
equations of motion coming from the given matter model? To say that the
extension is regular means that it satisfies certain differentiability
properties. What degree of differentiability is to be chosen? This is
not the place to give a general discussion of these questions - in the 
following they will only be answered insofar as this is directly relevant
to the main discussion.

When discussing strong cosmic censorship for solutions which are suitable
for being used as cosmological models there are two typical regimes which
come up. These are the approach to a singularity and an epoch of
unending expansion. This Letter is chiefly concerned with the second of
these regimes although some remarks will also be made on the first.
The approach to the singularity will be considered in detail in a 
related paper \cite{dr3}.

How can inextendibility be shown? A standard approach \cite{chrusciel}
is to show that either causal geodesics are complete or that, if they
are incomplete, some curvature invariant blows up along any incomplete
direction. Usually this requires detailed information about the
asymptotic behaviour of solutions of the Einstein-matter equations
in the given asymptotic regime and this is hard, even under strong
symmetry assumptions on the solutions considered. This paper presents
a method which in many cases is much simpler. It applies to spacetimes
with Killing vectors.
  
\section{Extensions of Killing vectors}

Let $X^\alpha$ be a Killing vector field on a Lorentzian manifold  
$(M,g_{\alpha\beta})$. It satisfies the well-known relation 
\begin{equation}\label{basic}
\nabla_\alpha\nabla_\beta X_\gamma=R_{\alpha\beta\gamma\delta}X^\delta.
\end{equation}
Next suppose that the original spacetime can be embedded isometrically
in another spacetime $(\tilde M,\tilde g_{\alpha\beta})$. Let $H$
be the boundary of the image of $M$ in $\tilde M$. Suppose further
that if $p\in H$ there is a local coordinate system $\{x^1,x^2,x^3,x^4\}$
such that $p$ is at the origin, the part of $H$ covered by these
coordinates coincides with the set defined by the equation 
$x^4=f(x^1,x^2,x^3)$ for some continuous function $f$ and the part
of $M$ covered coincides with the set defined by $x^4<f(x^1,x^2,x^3)$. We 
now think of the equation (\ref{basic}) as referring to components with
respect to these local coordinates.

It will be assumed that the metric $\tilde g_{\alpha\beta}$ is $C^2$,
so that the components of its curvature are continuous. Define 
$Y_{\alpha\beta}=\partial_\alpha X_\beta$. Then (\ref{basic}) implies,
by writing out the covariant derivatives explicitly in terms of Christoffel
symbols, the relation:
\begin{equation}\label{propagation}
\partial_4 Y_{\beta\gamma}=A^{\sigma\tau}_{\beta\gamma}Y_{\sigma\tau}
+B^\delta_{\beta\gamma}X_\delta
\end{equation}
for some continuous coefficients $A^{\sigma\tau}_{\beta\gamma}$ and 
$B^\delta$. Moreover $\partial_4 X_\alpha=Y_{4\alpha}$. Hence along
the coordinate lines of $x^4$ the collection $(X_\alpha, Y_{\alpha\beta})$
satisifies a linear system of ordinary differential equations with
coefficients which are continuous up to and including $H$. It follows 
that the components $X_\alpha$ and their first order partial derivatives 
are bounded as $H$ is approached. Hence these components are uniformly 
continuous and as a consequence they extend continuously to $H$ within the 
given coordinate system. To sum up, under the given assumptions, Killing 
vectors of $g_{\alpha\beta}$ have continuous local extensions to $H$.

\section{Symmetric spacetimes}

Consider first the case of a Lorentzian metric with $T^2$-symmetry. In this 
case it is assumed that that there are two commuting spacelike Killing 
vectors without fixed points. Call them $X$ and $Y$. It is possible to
choose local coordinates so that they take the form $\partial/\partial x$
and $\partial/\partial y$. Then the components of the metric in these
coordinates are independent of $x$ and $y$. It is common to define 
a quantity $R$ as a constant times the square root of the determinant of 
the metric induced on an orbit of the local group action generated by the 
Killing vectors. This can also be defined in a coordinate-invariant way
as the square root of the determinant of the matrix of inner products of
the Killing vectors $X$ and $Y$. With the latter definition it is clear
from the discussion of the last section that when the metric has a 
suitable $C^2$ extension the function $R$ extends continuously to the 
boundary $H$ of the extension. 

The information about $R$ just obtained will now be compared with the 
results of \cite{arw}. If $g_{\alpha\beta}$ is the metric defined 
by the maximal globally hyperbolic development of some Cauchy data
then the boundary $H$ of this spacetime in any extension is called the
Cauchy horizon. The boundary is known to satisfy the conditions required
in the previous section - the function $f$ can even be taken to be
Lipschitz continuous \cite{he}. Thus $R$ extends continuously 
to $H$. It follows from Theorem 2 of \cite{arw} that if the spacetime
concerned is a solution of the Einstein-Vlasov system which is initially 
expanding then $R$ tends to infinity along any future-directed causal curve. 
As a consequence, no future-directed causal curve can tend to a point of 
$H$ and no future extension of the original spacetime is possible. Thus 
the part of cosmic censorship referring to the future is settled for these 
spacetimes. To say it another way, it has been proved that the future
Cauchy horizon is empty. Note that this result in particular covers 
vacuum spacetimes with this symmetry. With the method of the present 
paper it follows immediately from \cite{bcim}. This result will now 
be formulated as a theorem:

\noindent
{\bf Theorem} The maximal globally hyperbolic development of data
for the Einstein-Vlasov equation with $T^2$ symmetry which are expanding 
cannot be extended in a $C^2$ way so as to allow an inextendible 
future-directed causal curve in the original spacetime to be continued.

The inextendibility result just derived was not previously known
for general $T^2$-symmetric vacuum spacetimes or for plane-symmetric
solutions of the Einstein-Vlasov equations, a very special case.
It was known for Gowdy spacetimes  but the proof was based on a
detailed determination of the asymptotics \cite{ringstrom} which was 
very complicated. Of course the asymptotics is in itself of great
interest but the point of this paper is to show that the question
of inextendibility can be handled with much simpler methods. 

Consider next the case of spherical or hyperbolic symmetry. In these
symmetry types there are three Killing vectors but these do not all
commute and they have fixed points. Hence the approach used for 
$T^2$-symmetry must be modified. To do this we use the fact that there
are three Killing vectors $X$, $Y$ and $Z$ on the standard sphere with
the property that the sum of the squares of their lengths is constant. 
This can be seen by considering the standard embedding of the unit 
sphere in $\R^3$ for which these Killing vectors can be taken as 
$x\partial_y-y\partial_x$, $y\partial_z-z\partial_y$ and
$z\partial_x-x\partial_z$. There are corresponding Killing vectors on
the spacetime and locally each of them extends continuously to $H$.
The sum of the squares of the lengths of $X$, $Y$ and $Z$ is
proportional to $R$ and so it can be concluded that $R$ extends 
continuously to $H$. It follows that a spherically symmetric spacetime which 
is a maximal globally hyperbolic development where $R$ goes to infinity along 
any future-directed causal geodesic cannot be extended to the future. In fact
spherically symmetric spacetimes tend to recollapse but in the presence of
a positive cosmological constant this result can be applied. In 
the case where it is known that $r$ tends to infinity geodesic
completeness has also been proved \cite{tchanou}. Nevertheless it
represents a simplified proof of inextendibility in that case and 
related applications are likely to come up in the future.

Spacetimes with hyperbolic symmetry can be handled in a similar way. This
time we choose Killing vectors $X$, $Y$ and $Z$ on hyperbolic space with the 
property that the combination $|X|^2+|Y|^2-|Z|^2$ is constant, where the 
modulus denotes the length with respect to the given metric. To see that 
vectors of this kind exist, consider the standard embedding of hyperbolic 
space in three-dimensional Minkowski space with metric $dx^2+dy^2-dz^2$
as the hyperboloid $x^2+y^2-z^2=-1$ and take the vectors 
$x\partial _z+z\partial_x$, $y\partial_z+z\partial_y$ and
$x\d_y-y\d_x$. Now take the corresponding Killing vectors on spacetime
and put them into the quadratic form introduced above. This quantity is
proportional to $R$. Following the same procedure as above gives a 
criterion for inextendibility in this case too. Combining it with a result 
of \cite{arr} shows that solutions of the Einstein-Vlasov system with 
hyperbolic symmetry cannot be extended to the future. This was previously 
known only in a small data situation \cite{rein}.

Using the results of Tegankong \cite{tegankong} it can be shown that
maximal globally hyperbolic developments are inextendible in the future in
the case of solutions of the Einstein-Vlasov-scalar field system with
plane or hyperbolic symmetry. The methods used here can 
be extended to the case of spacetimes with only one spacelike Killing 
vector (spacetimes with $U(1)$ symmetry). In that case the analogue of 
$R$ is the length of the Killing vector. An inextendibility result
for vacuum spacetimes with small initial data follows from \cite{cbm}.
In fact in that case a lot more control on the asymptotics is available
and geodesic completeness can be proved \cite{cotsakis}.

\section{Concluding remarks}

Strong cosmic censorship is a hard problem and for this reason it is
common to study restricted problems for spacetimes with symmetry.
In this paper a powerful method has been presented for proving 
inextendibility of cosmological spacetimes with symmetry in the future.
No symmetry is assumed of possible extensions. Straightforward consequences 
include a number of new results on classes of spacetimes frequently 
studied in the literature together with major simplifications of proofs of 
known theorems.

It should be noted that the method presented here only involves
the geometry of an extension. Properties of matter fields play
no role. On the other hand the control of the quantity $R$ which is 
necessary in order for the method to be applicable in any given example
relies on a previous analysis of the coupled Einstein-matter equations.

The results which have been presented here shift the emphasis in
proving cosmic censorship to the analysis of the initial singularity.
For that case the procedure of this paper says that at any potential
singularity on the boundary of the maximal Cauchy development is
such that the Killing vectors extend continuously (for any of the 
symmetry types considered here), as does the function $R$. It is
known in certain cases by other arguments that the extension
of $R$ is constant on the boundary. The limiting constant value may be 
positive or zero. The further insights which may be obtained in this way 
on the structure of singularities will be discussed elsewhere \cite{dr3}.

\vskip 10pt\noindent
{\bf Acknowledgment} ADR thanks the Isaac Newton Institute, Cambridge, for
hospitality while this work was being done.

\end{document}